\documentclass[12pt,a4paper]{article}

\usepackage[utf8]{inputenc}
\usepackage[T1]{fontenc}
\usepackage[english]{babel}

\usepackage{csquotes}

\usepackage{lmodern}
\usepackage{microtype}

\usepackage[a4paper,margin=2.5cm]{geometry}
\usepackage{setspace}
\usepackage{parskip}

\usepackage{graphicx}
\usepackage{booktabs}
\usepackage{caption}
\usepackage{enumitem}
\setlist{topsep=2pt,itemsep=2pt,parsep=2pt}

\usepackage{titlesec}
\titleformat{\section}{\large\bfseries}{\thesection}{0.6em}{}
\titleformat{\subsection}{\normalsize\bfseries}{\thesubsection}{0.6em}{}
\titlespacing*{\section}{0pt}{*1.2}{*0.6}
\titlespacing*{\subsection}{0pt}{*1.0}{*0.4}

\usepackage{fancyhdr}
\usepackage[hidelinks]{hyperref}
\hypersetup{
    pdftitle={Gibbs Article — European Diffusion},
    pdfauthor={Hector Giacomini},
    pdfcreator={LaTeX},
    pdfsubject={History of Science},
    pdfkeywords={Gibbs, statistical mechanics, European libraries}
}

\title{The Rapid Arrival of Josiah Willard Gibbs's \emph{Elementary Principles in Statistical Mechanics} in European University Libraries}
\author{Hector Giacomini\\ \small Institut Denis Poisson. Université d'Orléans --- Université de Tours --- CNRS.\\ 37200 Tours, France.\\ \texttt{giacominihector@gmail.com}}
\date{}  

\begin{document}
\maketitle

\begin{abstract}
This note offers an overview of how Josiah Willard Gibbs's \emph{Elementary Principles in Statistical Mechanics}, published simultaneously in London and New York in 1902, spread through European university libraries. Contrary to the received idea that the circulation of this text was slow, information gathered through direct contacts with numerous academic libraries, together with an examination of Yale University's archives, reveals an unexpectedly rapid material diffusion beginning on 15 March 1902. This early propagation can be explained by several channels: presentation copies sent by Yale University to leading universities, personal mailings by Gibbs himself to prominent scientists, and the distribution of copies by the American publisher to major scientific journals.
\end{abstract}

\section{Method and sources}
Factual data---such as the presence of copies, accession dates, and modes of acquisition---were obtained via email exchanges with academic library staff. In response to inquiries, these professionals consulted their inventories and historical registers with remarkable speed and professionalism. In some cases, archivists even sent photographs of register pages as well as of the library's copy of Gibbs's book.

\section{A brief biographical sketch of Josiah Willard Gibbs}
Josiah Willard Gibbs (1839--1903) was born in New Haven into a scholarly family \cite{wheeler1951, rukeyser1942, Klein1972, Bumstead1928}. His father was a professor of sacred literature at Yale University. After studies that combined science and the classical humanities---notably Latin and Greek---he completed in 1863 one of the earliest American doctorates, in mechanical engineering.

Thanks to a traveling fellowship, Gibbs spent 1866 to 1869 in Europe, moving in the scientific circles of Paris, Berlin, and Heidelberg. He attended courses by Helmholtz, Kirchhoff, and Clausius. This experience was formative, but Gibbs, reserved by temperament and little inclined to socializing, remained partly on the margins of European scientific networks.

Upon returning to Yale, he taught for several years without pay, in modest circumstances, before being appointed professor of mathematical physics in 1871. He spent the rest of his life there, never seeking to leave his university or to multiply publications. He died in 1903, a few months after the release of his major treatise. Unmarried, leading a sober life devoted entirely to research, Gibbs remained unknown to the general public. His scientific reputation grew slowly and quietly, culminating in the posthumous edition of his works in 1906 in two volumes: the first devoted to thermodynamics and statistical mechanics, the second to electromagnetism, optics, and various mathematical contributions, including vector calculus \cite{gibbs1906}.

Curiously, the Gibbs phenomenon---a persistent oscillation near discontinuities in Fourier series---is not discussed in the second volume. Observed experimentally by Albert Michelson around 1898, it was later explained mathematically by Gibbs, who showed that these oscillations do not disappear even as the number of terms in the series increases. This explanation was never published in an article, but circulated as a note among mathematicians of the time.

Including the vector-calculus textbook compiled from his lectures by his student Wilson \cite{wilson1901}, Gibbs published only five scientific works in his lifetime. Each, however, stands out for its exceptional depth and influence. He worked with methodical slowness and extreme rigor, publishing only when his ideas were fully matured. His biographers report that he spent nearly twenty years developing his results in statistical mechanics and reflecting on how best to present them. His treatise appeared just in time, a few months before his death.

Remarkably modest and little concerned with public recognition, Gibbs declined several prestigious, better-paid posts at institutions more renowned than Yale at the time. Apart from his European sojourn, he studied, taught, and conducted research at Yale throughout his career.

Gibbs explicitly acknowledged the influence of Ludwig Boltzmann. From the opening pages of his treatise, he situates his work in the continuity of the efforts of Clausius, Maxwell, and Boltzmann to ground thermodynamics on mechanical principles. By recognizing, clearly and already in the introduction, the contributions of his predecessors, Gibbs embodies the grandeur of the true scientific spirit---one that understands that knowledge is never the work of a single genius, but the fruit of a dialogue among researchers across generations.

\section{Some brief considerations on Gibbs's scientific work}
Josiah Willard Gibbs was one of the founders of modern thermodynamics and statistical mechanics. His monumental memoir, \emph{On the Equilibrium of Heterogeneous Substances}, published in two parts (1876 and 1878), established the foundations of modern chemical thermodynamics and revolutionized the discipline \cite{gibbs1876}. These works introduce the notion of chemical potential, systematize the use of potential functions---or free energies---as well as Legendre transforms. Their influence has endured in chemistry, physics, and biology.

Gibbs also formulated the famous phase rule, essential for the study of phase transitions. This rule determines the number of degrees of freedom of a thermodynamic system at equilibrium; that is, the number of intensive parameters (temperature, pressure, concentration, etc.) that can be adjusted independently without changing the number of phases present. It is widely used in materials physics, metallurgy, industrial chemistry, and the study of certain biological processes.

In 1901, his only doctoral student, Edwin Bidwell Wilson, published \emph{Vector Analysis}, based on Gibbs's lectures---as indicated by the book's subtitle. This treatise established the modern vector language in physics, greatly simplifying the writing of equations, notably Maxwell's. The book was part of the package sent by Yale to European universities as gifts. Thanks to Wilson, Gibbs's unpublished work on vector calculus was widely disseminated in Europe at the beginning of the twentieth century.

In 1902, Gibbs published his only book, \emph{Elementary Principles in Statistical Mechanics}, in the series of \emph{Yale Bicentennial Publications} \cite{gibbs1902}. This foundational text presents, in a systematic way, the method of ensembles---microcanonical, canonical, and grand canonical. Rather than reasoning about a single system, Gibbs proposes to study a conceptual collection of identical systems subject to the same constraints, and to deduce their average properties using probability theory. This approach allows one to recover macroscopic thermodynamic properties from microscopic constituents and the laws of mechanics.

This book is regarded as the official birth certificate of equilibrium statistical mechanics. Practically all the great treatises of the twentieth century drew inspiration from it, and it became a cornerstone of university teaching worldwide. It is a work of exceptional rigor and depth, which has influenced generations of physicists, mathematicians, and chemists.

Gibbs's method, born within the classical framework of thermodynamics, proved remarkably flexible, to the point of adapting to quantum physics. Its power lies in its generality: it rests on deep mathematical principles that transcend classical and quantum frameworks. Today, it is used in quantum statistical physics, in numerical simulation (via Gibbs sampling), in machine learning (probabilistic Bayesian models), and in atomistic modeling of proteins or membranes.

Gibbs offers a rigorous approach for linking the macroscopic laws of thermodynamics to the microscopic behavior of particles, without having to track each particle individually---unlike Ludwig Boltzmann and his kinetic theory of gases. Gibbs assumes thermodynamic equilibrium and deduces macroscopic properties by calculating probabilities over his abstract ensembles.

His approach is simpler, more rigorous, and more direct than Boltzmann's, whose kinetic method is based on a formidable nonlinear integro-differential equation. Boltzmann considers a very large number of particles (on the order of Avogadro's number), evolving in a phase space defined by their positions and momenta, according to the laws of Newtonian mechanics. Each configuration constitutes a microstate, and Boltzmann investigates the dynamical evolution of a concrete system toward equilibrium.

Gibbs, by contrast, does not seek to describe this temporal evolution. He assumes equilibrium has been reached and analyzes its properties by means of the probability density in phase space. The only fundamental properties required by his method---within Hamiltonian mechanics---are the conservation of total energy and that of volume in phase space, ensured by Liouville's theorem. This abstraction gives his formalism great flexibility, notably for adaptation to quantum physics.

As mentioned, Gibbs's formalism is limited to systems in thermal equilibrium, whereas Boltzmann's method makes it possible to study non-equilibrium systems, such as diffusion, transport phenomena, or viscosity. In Boltzmann's time, it was almost impossible to obtain concrete results from his kinetic equation. Only from the second half of the twentieth century did powerful mathematical tools allow progress in this area. Boltzmann therefore had to restrict himself to the study of very dilute gases.

For readers wishing to delve deeper into Ludwig Boltzmann's scientific work, Olivier Darrigol's book \cite{Darrigol2018} is an invaluable reference. Darrigol does more than provide a historical overview: he analyzes in depth Boltzmann's often complex writings, extracting their internal logic and theoretical intentions, while translating them into modern mathematical notation. He thus makes these texts accessible to physicists, philosophers, and historians of science. He also compares the approaches of Gibbs and Boltzmann, emphasizing that Gibbs proposes a more abstract and mathematically coherent method, founded on the probability density in phase space, in contrast with Boltzmann's more intuitive style. For a broader survey of Gibbs's scientific oeuvre, the reader may consult \cite{Mehra1998, brush1983, murthy2007}.

\section{The arrival of Gibbs's book in Europe}
Historiography has long maintained that Josiah Willard Gibbs's ideas spread very slowly in Europe. His writings, reputed to be dense and abstract, would have elicited little immediate response, and his 1902 book would have been largely ignored until the following decade. A recent article examines the impact of Gibbs's book during the first decade following its publication \cite{Popp2024}.

However, mapping the arrivals of this work in European university libraries reveals a markedly different reality. On the material level, diffusion was surprisingly rapid: copies were already available by mid-March 1902 in several major institutions. This does not guarantee either their reading or immediate assimilation, but it does testify to an effective and near-simultaneous institutional circulation on the continent. This early diffusion is explained by Yale University's initiative, which, to mark its bicentennial, sent presentation copies to many universities, as well as by Gibbs's personal mailings to scientific colleagues.

The following list results from a survey conducted among a wide range of academic libraries. Staff collaborated with notable responsiveness and professionalism. In some cases, photographs of accession registers and of the copies themselves were provided, offering valuable documentation for reconstructing the material trajectory of the book.

\section{Documented arrivals of Gibbs's book in European libraries}
\textbf{University of Oxford}: 1 April 1902, gift from Yale University.

\textbf{University of Basel}: 10 April 1902, gift from Yale University.

\textbf{University of Göttingen}: 15 April 1902, gift from Yale University.

\textbf{University of Heidelberg}: 15 April 1902, gift from Yale University.

\textbf{University of Erlangen--Nuremberg}: 31 July 1902, gift from Yale University.

\textbf{University of Strasbourg}: 15 April 1902, gift from Yale University.

\textbf{University of Vienna}: 19 April 1902, gift from Yale University.

\textbf{University of Turin --- Mathematics Library}: 20 January 1903, purchase via a Turin bookshop.

\textbf{University of Bologna --- Mathematics Library}: 1902, very likely a gift from Yale University; precise arrival date unavailable.

\textbf{University of Padua}: 30 October 1902, purchase via a Padua bookshop.

\textbf{Scuola Normale Superiore di Pisa}: 1902 edition---gift received in 2002 from Charlotte Janice Brudno, wife of the mathematician and historian of science Clifford Truesdell---the copy bears a stamp of Marcel Brillouin and Yale University's gold seal.

\textbf{University of Pisa --- Library of Mathematics, Computer Science and Physics}: 1902 edition with Yale University's gold seal---donated by Professor Pierluigi Riani in 2014.

\textbf{University of Cagliari --- Biomedical and Scientific Library}: 1902, very likely a gift from Yale University; precise arrival date unavailable.

\textbf{University of Leipzig}: 1902---archives destroyed in fires caused by Second World War bombings---very likely a gift from Yale University.

\textbf{Technical University of Munich}: 1902, archives unavailable---very likely a gift from Yale University.

\textbf{University of Berlin}: 1902, gift from Yale University via subscription to the \emph{Yale Bicentennial Publications} series; archives unavailable.

\textbf{École Polytechnique (Paris)}: 1913---purchase---1902 edition.

\textbf{Rijksmuseum Boerhaave Library of Natural Sciences (Leiden)}: 1902 edition transferred to the museum by the Lorentz Institute---this copy is annotated by the physicist Paul Ehrenfest.

\textbf{British Library}: 1902 edition---transferred by the Sir John Cass Technical Institute in 1983---very likely a gift from Yale University.

\textbf{University College London (UCL)}: 1902 edition---transferred by the Finsbury Technical College in 1922---very likely a gift from Yale University.

The copies sent by Yale to university libraries, as well as those personally addressed by Gibbs to leading scientists, were generally bound in the publisher’s standard blue cloth with the Yale seal stamped in gilt — the same binding that was also distributed commercially by Charles Scribner’s Sons (New York) and Edward Arnold (London).

\section{The Rijksmuseum Boerhaave (Leiden) copy annotated by Paul Ehrenfest}
One singular case deserves emphasis: the Rijksmuseum Boerhaave, a museum of the natural sciences in Leiden, preserves an annotated copy of the 1902 first edition of Gibbs's treatise bearing Paul Ehrenfest's handwritten marks. Long kept in the library of the Lorentz Institute---founded in 1921---this volume entered the museum's collections only in 1966.

The historical importance of this copy lies in Ehrenfest's central role in transmitting statistical mechanics. A student of Ludwig Boltzmann at the University of Vienna, he defended his thesis in 1904 on Hertzian mechanics and the motion of rigid bodies in fluids. In 1912, he succeeded Hendrik Antoon Lorentz in the chair of theoretical physics at the University of Leiden, a position he held until his death in 1933.

Ehrenfest was one of the great conduits of Boltzmann's thought to the rising generation of physicists. Discovering his annotations in Gibbs's book provides concrete, valuable evidence of the work's immediate reception within Boltzmann's intellectual circle. The rediscovery of this copy in a museum, outside the traditional circuits of university libraries, is remarkable in itself.

\section{The British Library copy}
Another copy merits particular attention: the one currently held at the British Library, whose origin traces back to the library of the Sir John Cass Technical Institute. Founded in 1899 thanks to the financial support of the John Cass Foundation---a philanthropic enterprise established by John Cass, an eighteenth-century politician and patron---the institute welcomed its first students in 1902. It offered technical and vocational education primarily aimed at young adults and adolescents from London's working-class milieus. In 1950, the institution was renamed the John Cass College, and its curriculum broadened beyond strictly technical subjects.

The copy in question still bears Yale University's stamps. The exact path of this volume remains enigmatic: how did a book probably issued from Yale University's official mailing end up not at a major university but at a tertiary-level technical institution in London?

Long on the margins of traditional academic circuits, this volume entered the British Library's collections only in 1983. It thus illustrates an occasionally unexpected trajectory of Gibbs's treatise in Europe, revealing that the material circulation of scientific works can follow indirect routes that elude standard institutional logics.

\section{The University College London copy}
As in the previous case, this copy was transferred to a prestigious institution from a London technical school, the Finsbury Technical College, created in 1893. The school aimed to be a model institution for the trades, intended for artisans and those preparing for intermediate industrial positions. It trained about 7{,}000 engineers and technicians and served as a model for future technical colleges across the United Kingdom until 1926.

\section{The Scuola Normale Superiore di Pisa copy}
This copy has a particularly rich history. It appears to have belonged to Marcel Brillouin, a professor at the Collège de France in 1902. A stamp bearing his name appears in the book, attesting to this provenance. Brillouin held dual doctorates in physics and mathematics, was a former student of the École Normale Supérieure in Paris, and was elected to the Académie des Sciences in 1921. His scientific work covered a wide range of fields: hydrodynamics, electromagnetic induction, wave mechanics, geophysics, and the plasticity of solids. He was also among the participants at the first Solvay Conference (1911), alongside the leading figures of physics of the day.
The book was later acquired by Clifford Truesdell, one of the most singular and erudite figures of the twentieth century. Both a mathematician and a historian of science, Truesdell was born in Los Angeles in 1919 and earned his doctorate in mathematics at Princeton in 1943. Before that, he spent two years in Europe studying Latin, ancient Greek, French, German, and Italian---languages he would use in his historical research.

Considered the modern founder of continuum mechanics, he developed with Walter Noll a rigorous formulation of rational mechanics, aiming to clarify the mathematical foundations of thermodynamics and fluid mechanics. Truesdell was also passionate about the history of mathematics, especially Leonhard Euler, whose collected works he edited in several volumes. He founded two major journals: \emph{Archive for Rational Mechanics and Analysis} and \emph{Archive for History of Exact Sciences}.

In 1978, Truesdell received the \emph{Ordine del Cherubino}, an honorary distinction awarded by the University of Pisa to individuals who have made exceptional contributions to the institution's intellectual and scientific life. This honor reflects the esteem in which he was held by the Italian scientific community, notably for his work in rational mechanics and the history of science.

He maintained close ties with many Italian researchers, and his writings were widely read and translated in Italy. He was perfectly fluent in Italian and sometimes published in that language.

Two years after his death, his widow, Charlotte Janice Brudno, donated this copy to the Scuola Normale Superiore di Pisa.

\section{The University of Basel copy}
The University Library of Basel was the first institution on the European continent to receive Gibbs’ \emph{Elementary Principles in Statistical Mechanics}. The book arrived on 10 April 1902 as part of a donation of twenty-three Bicentennial Publications from Yale University. Founded in 1460, the University of Basel is the oldest university in Switzerland. Its geographical position is particularly notable: located at a triple frontier between Switzerland, Germany, and France, the city has long been a hub of transportation and academic exchange. The straight-line distances to nearby university cities are approximately:

Freiburg im Breisgau, Germany – 52 km

Bern, Switzerland – 69 km

Zurich, Switzerland – 76 km

Strasbourg, France – 113 km

\section{On the difficulty of reading Gibbs's book}
Gibbs's book is renowned for its conceptual density and demanding mathematical style. It is often considered a difficult read even for experienced physicists. Gibbs adopts a formal, rigorous language, without attempting to popularize his ideas. He presupposes in the reader a prior mastery of the foundations of Hamiltonian mechanics and probability theory. Seldom illustrated by intuitive physical examples, his demonstrations favor an axiomatic, abstract, and methodically pared-down approach.

In this treatise, Gibbs introduces the microcanonical, canonical, and grand-canonical ensembles---now fundamental in statistical mechanics---which at the time appeared as particularly abstract theoretical constructions. European physicists, accustomed to Boltzmann's more intuitive formalism, faced a radically different style. The notations and reasoning employed by Gibbs, matured over years of solitary reflection, were unusual and sometimes disconcerting to his contemporaries.

Henri Poincaré explicitly mentioned Gibbs's book at the \emph{International Congress of Arts and Sciences} in St. Louis in 1904, in his lecture entitled \emph{The Present and Future of Mathematical Physics}. He recognized the theoretical importance of the work while noting its difficulty of access, due to its abstraction and advanced mathematical formalism.

For physicists on the European continent, the language barrier presented an additional obstacle: mastery of English was not yet widespread in scientific circles.

That said, a pragmatic reader could bypass some difficulties by focusing on the essential sections. The first forty pages of the book are devoted mainly to Hamiltonian formalism, to epistemological reflections on the nature of statistical laws, and to philosophical considerations on mechanics. Though profound, these sections are not indispensable for assimilating the concrete tools of statistical mechanics. Likewise, the long developments on thermodynamics can be skipped by readers already familiar with the subject.

Chapters 4 to 9 contain the genuinely novel results essential to understanding statistical mechanics. It is in these pages that Gibbs sets out the foundations of his ensemble method, which would become one of the pillars of twentieth-century theoretical physics.

\section{The copy sent by Gibbs to Henri Poincaré}
An inscribed copy of \emph{Elementary Principles in Statistical Mechanics}, sent by Gibbs to Henri Poincaré---very likely in March 1902---constitutes an important historical document. It bears witness to the mutual intellectual respect between two major figures of turn-of-the-century science, embodying the dialogue between American and French scientific traditions.

This volume resurfaced in prestigious auction sales, notably at \emph{Sotheby's} and \emph{Christie's}, where it attracted the interest of collectors specializing in scientific manuscripts. Its value rests as much on the rarity of the original printing as on the symbolic import of the inscription, which links two schools of thought at a pivotal moment in the history of physics.

The most recent known appearance of this copy dates to May 2019, when it was offered for sale by \emph{Sophia Rare Books}. Listed in their ``recent arrivals'' catalogue, it was described as a first issue accompanied by a handwritten dedication from the author to Jules-Henri Poincaré, as follows:

``\emph{Original edition, presentation copy to the great French mathematician and physicist Henri Poincaré}.''

The asking price was US\$17{,}500, with the clarification that this was a fixed-price offer and not an auction. The description emphasized the rarity and exceptional scientific value of the work, regarded as a foundational classic of statistical mechanics.

For reasons of historical and patrimonial coherence, it would be fitting for this volume to join the collections of the library of the Institut Henri Poincaré in Paris, where it could be preserved as a material witness to a remarkable intellectual exchange.

\section{The copy sent by Gibbs to John William Strutt, Lord Rayleigh}
John William Strutt, better known by his title Lord Rayleigh, was a leading British physicist and the 1904 Nobel laureate in physics. He is notably famous for explaining why the sky is blue, by studying the scattering of light by air molecules---a phenomenon now known as Rayleigh scattering. He also identified a type of surface seismic wave that bears his name and contributed to the precise definition of electrical units during his tenure at the University of Cambridge.

Together with William Ramsay, he discovered argon in 1894, a major advance that reshaped the chemistry of gases and enabled the identification of the noble-gas family. In 1879, he succeeded James Clerk Maxwell as head of the Cavendish Laboratory, one of the most prestigious physics research centers. He was also president of the Royal Society from 1905 to 1908, then chancellor of the University of Cambridge from 1908 to 1919.

An inscribed copy addressed to Lord Rayleigh was auctioned in New York on 29 October 1998 by \emph{Christie's}.

\section{Copies sent by Gibbs to various leading European scientists}
At the beginning of 1902, Gibbs undertook a broad campaign to disseminate his book by personally sending copies to a wide array of scientists---physicists, mathematicians, chemists, and astronomers---as well as to several learned institutions in Europe. This gesture testifies to his desire to share his ideas with the leading intellectual figures of his time, despite his legendary discretion.

The following list enumerates the European recipients identified to date, based on the archives consulted. It reflects the breadth and diversity of the intellectual network Gibbs targeted and helps to clarify the channels through which his treatise circulated within the scientific landscape of the early twentieth century.

\textbf{France}:

Henri Poincaré---mathematician and physicist---Paris.

Pierre Duhem---theoretical physicist---Bordeaux.

Joseph Valentin Boussinesq---mathematician and physicist---Paris.

Jacques Hadamard---mathematician---Paris.

Henry Le Chatelier---chemist---Paris.
 
Paul Émile Appell---mathematician---Paris.

Académie des Sciences de Paris.

\textbf{United Kingdom}:

John William Strutt, Lord Rayleigh---London---Nobel Prize in 1904.

George H. Darwin---astronomer---Cambridge.

Samuel H. Burbury---applied mathematician---Oxford.

William Thomson, Lord Kelvin---physicist---Glasgow.

James Dewar---chemist and experimental physicist---Cambridge.

J.~J. Thomson---physicist---Cambridge---Nobel Prize in 1906.

Sir Robert Ball---astronomer---Cambridge.

Joseph Larmor---physicist and mathematician---Cambridge.

Henry William Watson---applied mathematician---Oxford.

Royal Society (London).

\textbf{Germany}:

Felix Klein---mathematician---Göttingen.
 
Wilhelm Ostwald---chemist and physicist---Leipzig---Nobel Prize in 1909.

Max Planck---theoretical physicist---Berlin---Nobel Prize in 1918.

Friedrich Kohlrausch---experimental physicist---Berlin.

Friedrich Engel---mathematician---Leipzig.

Berlin Academy.

\textbf{Austria}:

Ludwig Boltzmann---theoretical physicist---Vienna.

\textbf{Netherlands}:

Hendrik A. Lorentz---theoretical physicist---Leiden---Nobel Prize in 1902.

Heike Kamerlingh Onnes---experimental physicist---Leiden---Nobel Prize in 1913.

Johannes D. van der Waals---physicist---Amsterdam---Nobel Prize in 1910.

Jacobus Henricus van 't Hoff---chemist---Amsterdam---Nobel Prize in 1901.

Hendrik W. Bakhuis Roozeboom---chemist---Amsterdam.

\textbf{Italy}:

Luigi Cremona---mathematician---Rome.

\textbf{Sweden}:

Gösta Mittag-Leffler---mathematician---Stockholm---founder of the renowned journal \emph{Acta Mathematica}.

This list appears in small print at the end of L.~P. Wheeler's biography of Gibbs \cite{wheeler1951}. It is found on pages 257 and 258, and, according to our online research, has gone largely unnoticed. Yale University's archives preserve a letter of thanks from Max Planck for the copy sent by Gibbs, received in New Haven on 29 March 1902, as well as another letter from Hendrik Lorentz, dated 1 April 1902, received at nearly the same time. These documents suggest that Gibbs's mailings were received around mid-March 1902.

\section{Copies sent by the American publisher to scientific journals and editors}

\emph{Nature}---London.

\emph{Philosophical Magazine}---London.

\emph{Fortschritte der Physik}---Braunschweig---Germany.

\emph{Jahrbuch über die Fortschritte der Mathematik}---sent to Emil Lampe---Berlin.

\emph{Beiblätter zu den Annalen der Physik}---sent to Walter König---Leipzig.

\emph{Zeitschrift für Physikalische Chemie}---sent to Wilhelm Ostwald---Leipzig.

\emph{Monatshefte für Mathematik und Physik}---sent to Gustav von Escherich---Vienna.

\emph{Physikalische Zeitschrift}---sent to Max Abraham---Göttingen---Germany.

\emph{Bulletin des Sciences Mathématiques}---sent to Gaston Darboux---Paris.

\emph{Il Nuovo Cimento}---sent to Angelo Batelli---Pisa.

\emph{L'Enseignement Mathématique}---sent to Charles-Ange Laisant---Paris.

\emph{Revue Semestrielle des Publications Mathématiques}---sent to Johannes Korteweg---Amsterdam.

\emph{Jahresbericht der Deutschen Mathematiker}---sent to A.~Gutzme---Halle---Germany.

\emph{Le Matematiche Pure ed Applicate}---sent to Cristoforo Alasia---Tempio---Italy.

This list was compiled by Gibbs himself and appears alongside the previous list in Wheeler's biography, likewise in small print and largely unnoticed.

\section{Conclusions}
In March and April 1902, the diffusion of Josiah Willard Gibbs's \emph{Elementary Principles in Statistical Mechanics} gained remarkable momentum in Europe. Three parallel channels opened almost simultaneously. European libraries received lots shipped by Yale University. Scientific journals received copies for review and announcement. A series of courtesy mailings by Gibbs himself reached the leading scientific figures and academies of Europe.

This triple mechanism---institutions, journals, scholars---explains why, by the summer of 1902, the work was not confined merely to a material presence in Europe: it was already integrated into the networks of review and announcement in the scientific press. The notion that the book was difficult to access thus loses considerable credibility.

\section{Acknowledgments}
I wish to express my gratitude to the library staff with whom I corresponded by email. Their availability, speed of replies, precision in research, and attention to the material details of books and registers were of great importance for this inquiry. The photographs provided, the patient verifications, and the clarity of explanations all testify to genuine dedication to their work, animated by a true love of the profession.

\end{document}